\documentclass[twocolumn,aps,prb,showpacs,preprintnumbers,amsmath,amssymb,psfrac]{revtex4}

\usepackage{epsfig}
\usepackage{graphicx}
\usepackage{dcolumn}
\usepackage{bm}

\renewcommand{\dag}{^{\dagger}}

\def\gapp{\lower.35em\hbox{$\stackrel{\textstyle>}{\sim}$}}
\def\lapp{\lower.35em\hbox{$\stackrel{\textstyle<}{\sim}$}}

\begin{document}
%

\title{
Assisted hopping and interaction effects in impurity models. }
\author{L. Borda$^1$ and F. Guinea$^2$}
\affiliation{$^1$ Department f\"ur Physik and CENS,
Ludwig-Maiximilians-Universit\"at M\"unchen Theresienstr. 37,
D-80333 Munich, Germany\\
$^2$Instituto de Ciencia de Materiales de Madrid, CSIC, Cantoblanco, E-28049 Madrid, Spain\\
}
\date{\today}
\begin{abstract}
We study, using Numerical Renormalization Group methods, the
generalization of the Anderson impurity model where the hopping
depends on the filling of the impurity. We show that the model,
for sufficiently large values of the assisted hopping term, shows
a regime where local pairing correlations are enhanced. These
correlations involve pairs fluctuating between on site and nearest
neighbor positions.
\end{abstract}
%
%
%
%
\maketitle
\section{Introduction.}
The formation of local moments is the simplest and most
extensively studied manifestation of strong electron-electron
repulsion in impurities in metals\cite{A61,K64,H97}. This
phenomenon can also be observed in mesoscopic systems coupled to
metallic leads\cite{Getal98,Cetal98}. In the opposite limit, when
the "impurity" becomes a metallic grain whose electronic structure
needs to be described by many narrowly spaced levels, the
electron-electron repulsion leads to Coulomb blockade\cite{AL91}.
The intermediate regime when the energy scales of interest, the
temperature, the strength of the Coulomb repulsion and the level
spacing are not too different in magnitude is not so well
understood. A similar situation arises in the study of strongly
correlated systems. The analogous of the Anderson impurity model
is the Hubbard model\cite{H63,K63} for extended systems, which
leads to a Mott transition and the formation of local moments at
half filling. The opposite limit, when many levels within the unit
cell have to be included, leads, generically, to ordinary Fermi
liquid behavior.

The simplest modifications of the extended Hubbard model taking
into account the multiplicity of levels in the ions in the unit
cell lead to assisted hopping terms\cite{HM91,H93}, which modify
substantially the phase diagram, and tend to favor
superconductivity. Analogously, if the internal degrees of freedom
of the metallic grain are included, beyond the constant
interaction term which leads to Coulomb blockade, one finds non
equilibrium effects\cite{UG91,Betal00} which suppress the Coulomb
blockade. These effects are similar to the formation of an
excitonic resonance found in the excitation of core
electrons\cite{ND69}. One can also extend the Anderson impurity
model to include the effects associated to the existence of many
orbitals. When the influence of these orbitals is described in
terms of effective interactions within the restricted Hilbert
space of the usual Anderson model, one indeed finds assisted
hopping terms\cite{G03}. The resulting model has been analyzed
using a mean field, BCS like decoupling of the interaction
term\cite{G03}, and by the flow equation method\cite{SG04} (for a
description of the method, see\cite{W94,W01}), which is well
suited to the analysis of impurity
models\cite{KM94,KM96,LW96,M97,KM97}. These works show a tendency
towards local pairing away from half filling, in agreement with
the studies of bulk systems\cite{HM91,H93}. We present here more
accurate calculations, using the Numerical Renormalization Group
Method\cite{W75}, which characterize the low
energy properties of the model 
in a numerically exact way.

We examine the effect of the assisted hopping term on the local
density of states and find that with increasing assisted hopping
amplitude the peak characteristic to the mixed valence regime of
the Anderson model gets broadened and shifted to negative
frequencies. This result is consistent with the fact that the
assisted term enhances the occupation of the local level, also
found in the calculations. To demonstrate the effect of the new
term on the formation of the local moment, we compute the local
spin spectral function and find that the spin susceptibility gets
suppressed as the assisted hopping term is turned on. This
suppression might be interpreted as the trace of the pairing. We
also check the pairing correlations in a more direct way and
calculate the related spectral functions. Our main result is that
--although the effect of the assisted term on the pairing
correlations on the local level could be explained qualitatively
by just considering the renormalization of the level-- the
so-called off diagonal correlations (pairing between the
$d$-electron and the conduction electrons located at the impurity
position) are also enhanced.

The paper is organized as follows: 
The model, and the method of calculation are described in the next
section. Then, the main results are presented. Section IV gives
the main conclusions of our work.

\section{The model and method of calculation.}
\subsection{The model}
We describe the effects of the internal degrees of freedom of the
impurity in terms of effective interactions defined within the
restricted Hilbert space of the standard Anderson
impurity\cite{G03}. Using perturbation theory, this approach is
justified when the typical level spacing within the impurity,
$\Delta$, is much smaller than value of the Coulomb interaction,
$U$. Then, it is easy to show that the Anderson model is recovered
in the limit $\Delta / U \rightarrow \infty$. We will study the
Hamiltonian:
\begin{eqnarray}
{\cal H} &= &{\cal H}_K + {\cal H}_{imp}
+ {\cal H}_{hyb} + {\cal H}_{assisted} \nonumber \\
{\cal H}_K &= &\sum_{k,s}\epsilon_k c_{k,s}\dag c_{k,s} \nonumber \\
{\cal H}_{imp} &= &
\epsilon_d n_d+U n_{d,+}n_{d,-} \nonumber \\
{\cal H}_{hyb} &= & \sum_{k,s} V \left( c_{k,s}\dag d_s+d_s\dag
c_{k,s} \right)
\nonumber \\
{\cal H}_{assisted} &= &\sum_{k,s}dV n_{d,s}\left( c_{k,-s} \dag
d_{-s}+d_{-s}\dag c_{k,-s}\right) \label{hamil}
\end{eqnarray}
where ${\cal H}_A = {\cal H}_K + {\cal H}_{imp} + {\cal H}_{hyb}$
is the Anderson Hamiltonian, and the assisted hopping terms are
included in ${\cal H}_{assisted}$. We have also defined
$n_{d,s}=d_s\dag d_s$ and $n_d=n_{d,+}+n_{d,-}$. The parameter
$dV$ determines the strength of the assisted hopping effects.
For the sake of simplicity we consider a flat band with
constant density of states for the conduction electrons, with half
bandwidth $D$. In the following we use $D$ as the energy unit.

\subsection{The Numerical Renormalization Group Method.}

To compute different quantities numerically
we use Wilson's Numerical
Renormalization Group method\cite{W75,C99}.
In this method --after the logarithmic
discretization of the conduction band -- one maps the original
impurity Hamiltonian onto a semi-infinite chain with the impurity at the end.
One can show that as a consequence of the
logarithmic discretization the hopping along the
chain decreases exponentially, $t_n\sim\Lambda^{-n/2}$ where $\Lambda$ is
the discretization parameter, and $n$ is the index of the site in the chain.
The separation of the energy scales due to the decreasing hopping provides
the possibility to diagonalize the chain Hamiltonian iteratively and keep the
states with the lowest lying energy eigenvalues as most relevant ones. Since
we know the energy eigenstates and eigenvalues, we can calculate
thermodynamical and
dynamical quantities directly (e.g. spectral functions using their
Lehman representation).

\section{Results}
\begin{figure}
\includegraphics[width=8cm]{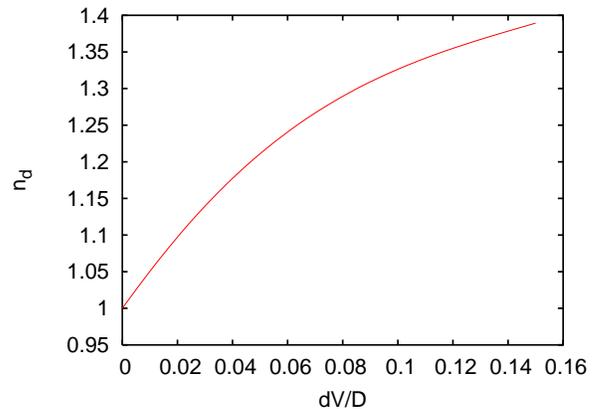}
  \begin{center}
    \caption{Value of $n_d$ as function of $dV$ for
    $V=0.2 , \epsilon_d = 0$ and $U = 0.05$.}
    \label{n_d}
\end{center}
\end{figure}

The model has four parameters, $\epsilon_d , U , V$ and $dV$.
Previous studies\cite{G03,SG04} suggest that the most significant
deviations from the standard Anderson impurity model take place
when the filling of the $d$ orbital is such that $1.2 \le \langle
n_d \rangle \le 1.8$.
We focus our attention to that regime as well.
The filling is determined by $\epsilon_d$,
$U$
and also by $dV$. The mean field decoupling of the assisted hopping term
shows that the latter term contributes to the
renormalization
of the position of
the level. The value of $n_d$ as function of $dV$ is given in
Fig.[\ref{n_d}]. Since the assisted hopping term renormalizes
the impurity level downwards, that term
favors the occupancy of the impurity level.

\begin{figure}
\includegraphics[width=8cm]{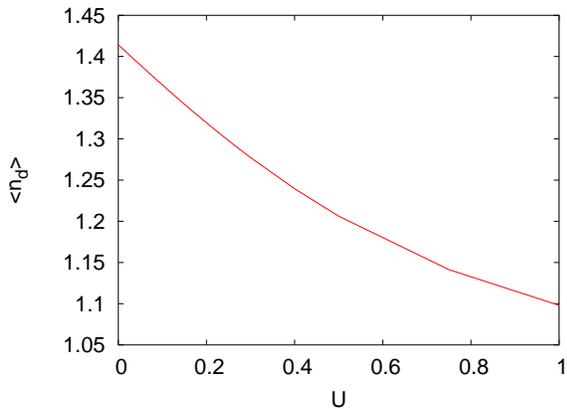}
  \begin{center}

    \caption{Value of $n_d$ as function of $U$ for
    $V=0.2 , \epsilon_d = 0$ and $dV = 0.15$.}
    \label{n_dU}
\end{center}
\end{figure}

The fact that the assisted hopping term, $dV$,
and the electron-electron repulsion $U$ have opposite effects can
be appreciated in Fig.[\ref{n_dU}] where the influence of the
value of $U$ on $n_d$, for a fixed value of $dV$, is shown.

\begin{figure}
\includegraphics[width=8cm]{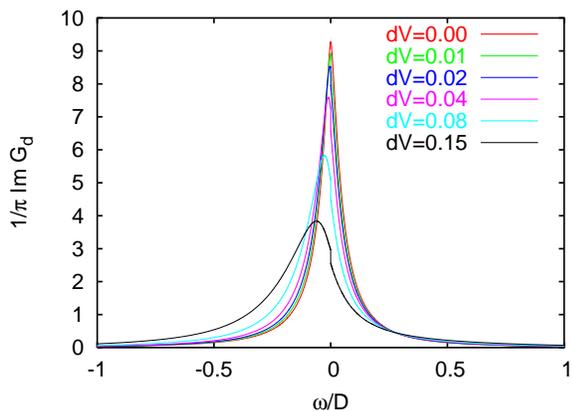}
  \begin{center}
    \caption{Imaginary part of the impurity one electron Green's function, $1/\pi\;G_d ( \omega )$, for
    $V=0.2 , \epsilon_d = 0$ and $U = 0.05$.}
    \label{Gd}
\end{center}
\end{figure}
\begin{figure}
\includegraphics[width=8cm]{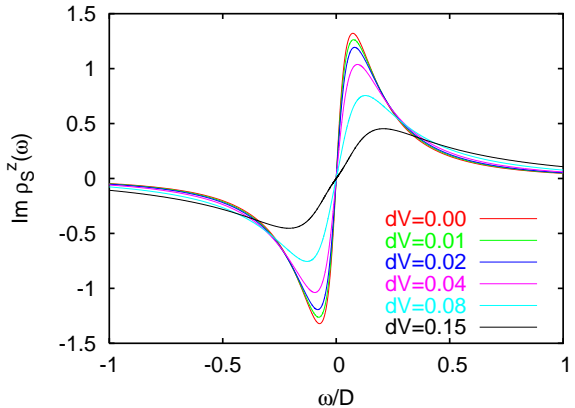}
  \begin{center}
    \caption{Spin-spin correlation function for $V=0.2 , \epsilon_d = 0$ and $U = 0.05$
    and different values of $dV$.}
    \label{Sz}
\end{center}
\end{figure}

Fig.[\ref{Gd}] shows the
imaginary part of the
one particle Green's function associated
to the localized level, defined as:
\begin{eqnarray}
\frac{1}{\pi} G_d ( \omega ) &=& \sum_{m , \sigma} \left| \langle
0 | d^{\dag}_\sigma | m \rangle \right|^2 \delta ( \omega -
\epsilon_m
)\nonumber\\
&+& \sum_{m , \sigma} \left| \langle 0 | d_\sigma | m \rangle
\right|^2 \delta ( \omega + \epsilon_m )\;,
\end{eqnarray}
where $|0\rangle$ ($|m\rangle$) is the ground ($m$th excited) state
of the full Hamiltonian with energy $\epsilon_0=0$ ($\epsilon_m$).

In presence of no assisted hopping, the local density of states
exhibits a broad peak at the Fermi level which is characteristic
to the mixed valence regime of the Anderson impurity model. As
$dV$ is increased from zero, 
the results show a broadening of
the resonance, also consistent with the fact that, at the mean
field level, the assisted hopping term modifies the effective one
particle hopping. 
As one can see in Fig[~\ref{Gd}], the
resonance gets not only broadened but also shifted down to
negative energies. This fact could already be foreseen from the
result for the occupation, since the occupation (at $T=0$) is just
the integral of the density of states for negative frequencies.
This enhancement of the occupation must leave its trace on the
local moment formation as well. 

The influence of the assisted hopping term on the formation of a
local moment can be inferred from Fig.[\ref{Sz}], which displays
the spin-spin correlation function:
\begin{equation}
{\rm Im}\; \varrho_S^z ( \omega ) = \frac{1}{2} \sum_n \left| \langle 0 | n_{d
\uparrow} - n_{d \downarrow} | n \rangle \right|^2 \delta ( \omega
- \epsilon_n )
\end{equation}
This function, for $dV = 0$ shows a pronounced peak at low
energies, which tend to the Kondo temperature below which the
local moment is quenched as $U$ increases. At higher energies,
${\rm Im}\; \varrho_S^z ( \omega ) \sim \omega^{-1}$, which corresponds to a Curie
susceptibility as function of temperature.
This behavior crosses over to a ${\rm Im}\; \varrho_S^z( \omega ) \sim \omega$
regime which is characteristic to the compensation of the moment.
(Alternatively, one could say that for very high frequencies
${\rm Im}\; \varrho_S^z ( \omega ) \sim \omega^{-1}$ as a consequence of the constant
time correlator of the spin for very short times, while for small
frequencies ${\rm Im}\; \varrho_S^z( \omega )$ shows a linear dependence on $\omega$
corresponding to the $\sim t^{-2}$ asymptotics of the correlation.
As $dV$ is turned on, this peak becomes
broader and it is shifted towards higher energies
indicating the suppression of the local moment.
This suppression is also consistent with the previous results and
might be considered as a fingerprint of the local pairing.

To get a more direct insight to the relation of the assisted hopping
and the pairing,
we now analyze the possible existence of pairing correlations in
the model. Within the reduced Hilbert space of the Anderson's
impurity model, we can define two types of such correlations:
\begin{widetext}
\begin{eqnarray}
F ( \omega ) &= &\sum_n \left| \langle 0 | d^{\dag}_\uparrow
d^{\dag}_\downarrow | n \rangle \right|^2 \delta ( \omega -
\epsilon_n ) + \left| \langle 0 | d_\uparrow d_\downarrow | n
\rangle \right|^2 \delta ( \omega + \epsilon_n ) \nonumber \\
F' ( \omega ) &= &\sum_n  \langle 0 | d^{\dag}_\uparrow
d^{\dag}_\downarrow | n \rangle \langle n | \left( d_\uparrow
c_\downarrow - d_\downarrow c_\uparrow ) | 0 \rangle \delta (
\omega - \epsilon_n \right) + \langle 0 | d_\uparrow d_\downarrow
| n \rangle \langle n | \left( d^{\dag}_\uparrow
c^{\dag}_\downarrow - d^{\dag}_\downarrow c^{\dag}_\uparrow
\right) | 0 \rangle \delta ( \omega + \epsilon_n ) \label{pairs}
\end{eqnarray}
\end{widetext}
where $c^{\dag}_\sigma = \sum_k c^{\dag}_{k \sigma}$, representing
the metal orbital closest to the impurity. The function $F (
\omega )$ in Eq.(\ref{pairs}) gives the magnitude of the on-site
pairing, as present, for instance, in the negative $U$ Anderson's
impurity model. The function $F' ( \omega )$ describes off
diagonal pairing. Virtual Cooper pairs resonate between the on-site
position and that in which one component of the pair is at the
impurity and the other is in the metal.

\begin{figure}
\includegraphics[width=8cm]{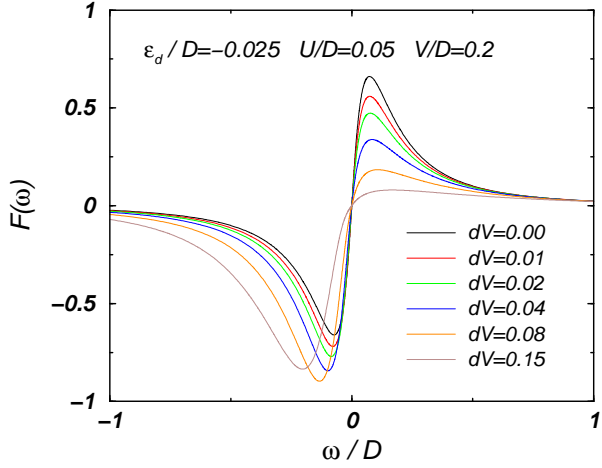}
  \begin{center}
    \caption{Local pair correlation, $F ( \omega )$ in Eq.(\protect\ref{pairs}) as
    function of $dV$ for $\epsilon = -0.025 , U = 0.05$ and $V = 0.2$.}
    \label{pair_onsite1}
\end{center}
\end{figure}

\begin{figure}
\includegraphics[width=8cm]{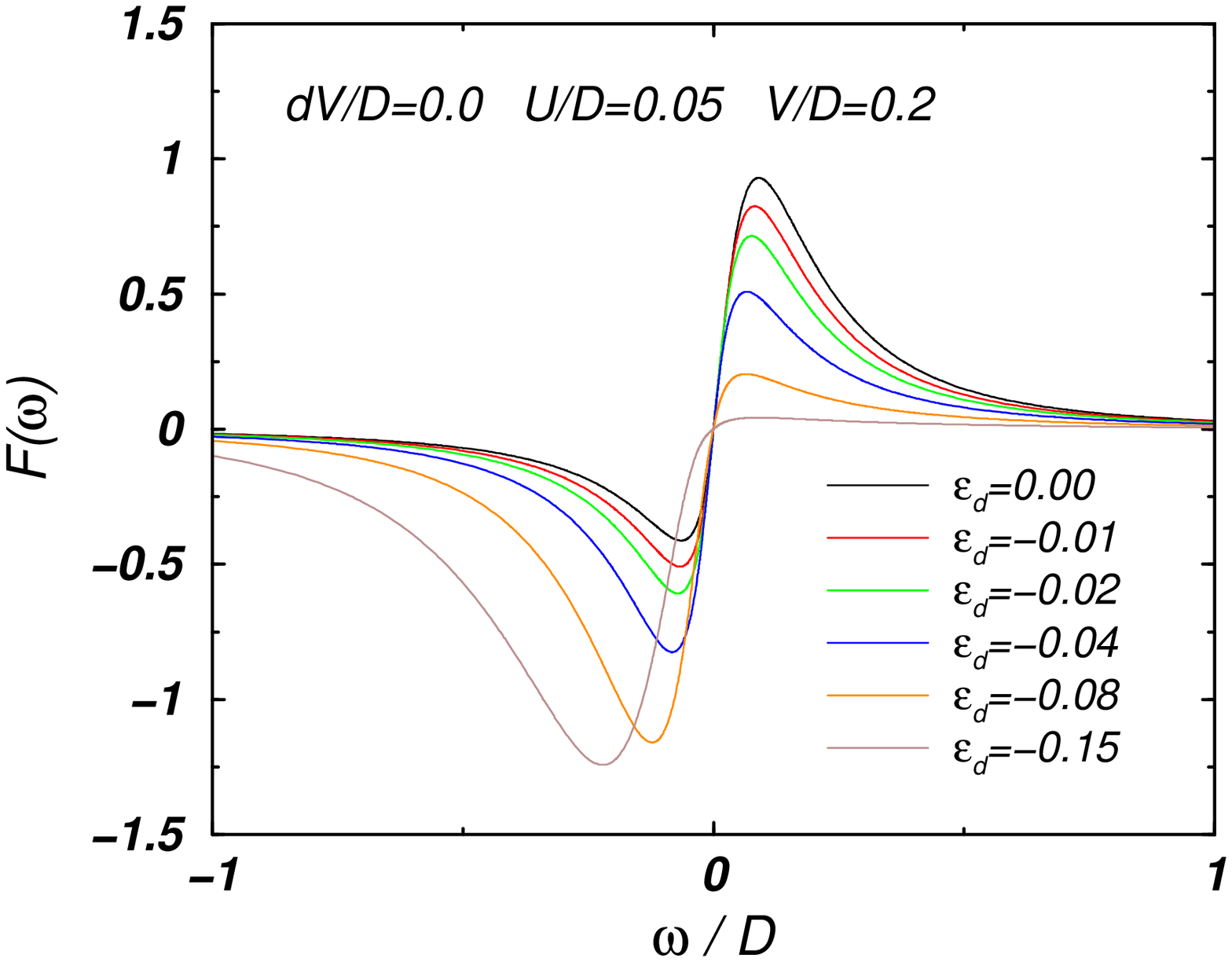}
  \begin{center}
    \caption{Local pair correlation, $F ( \omega )$ in Eq.(\protect\ref{pairs}) as
    function of $\epsilon$ for $dV = 0 , U = 0.05$ and $V = 0.2$.}
    \label{pair_onsite2}
\end{center}
\end{figure}
\begin{figure}
\includegraphics[width=8cm]{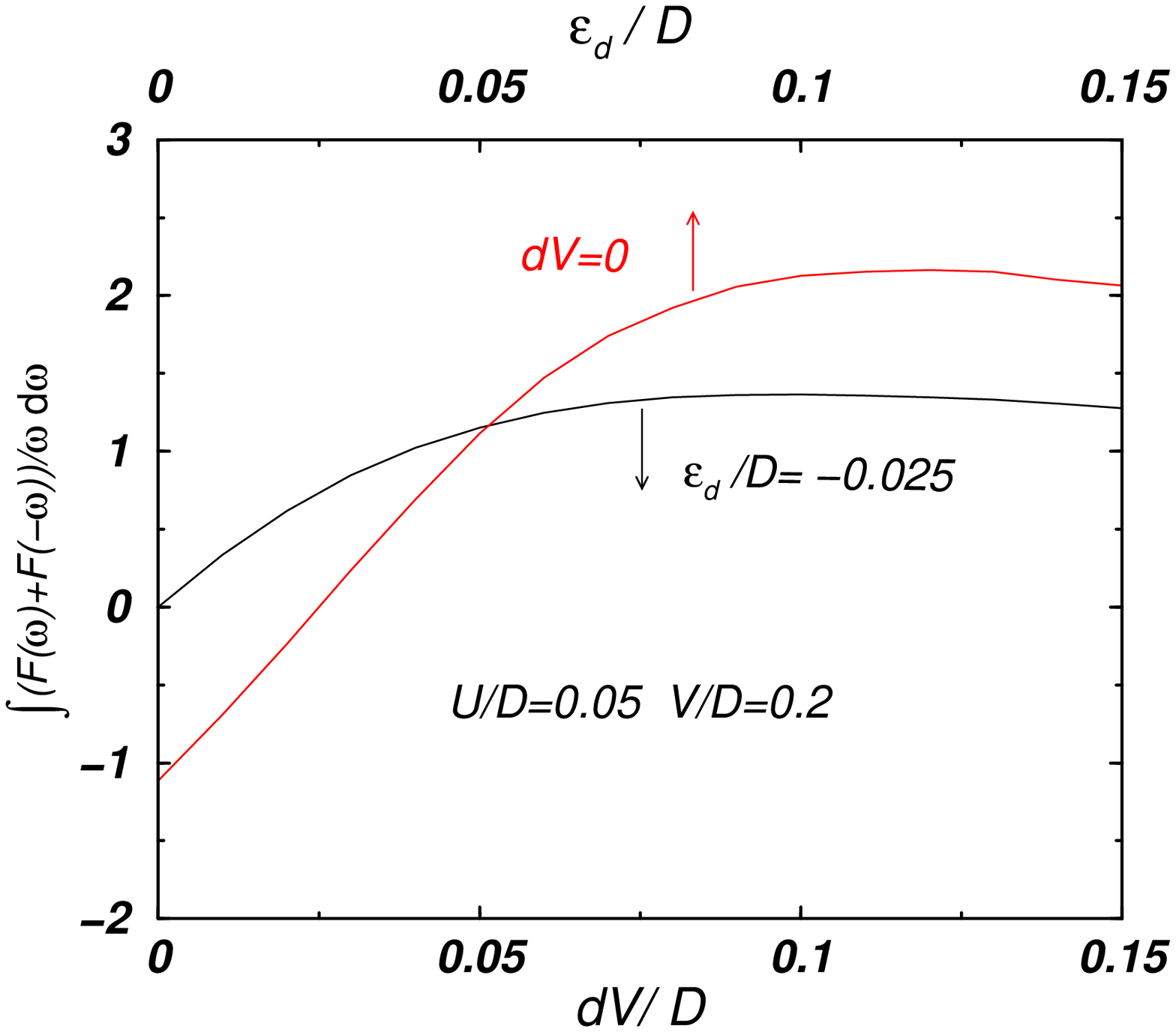}
  \begin{center}
    \caption{On site pair susceptibility, ${\cal F}$, as defined in Eq.(\protect\ref{susc})
    as function of $\epsilon$ for $dV = 0$ (top), and as function of $dV$ for $\epsilon =
    -0.025$
    (bottom).}
    \label{pair1}
\end{center}
\end{figure}

\begin{figure}
\includegraphics[width=8cm]{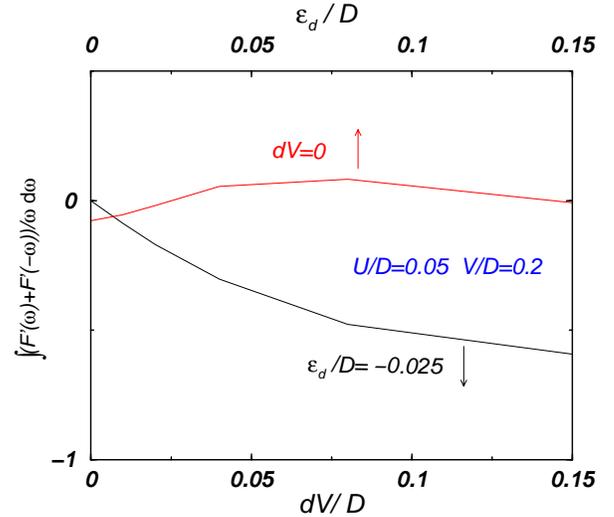}
  \begin{center}
    \caption{Off diagonal pair susceptibility, ${\cal F}'$, as defined in Eq.(\protect\ref{susc})
    as function of $\epsilon$ for $dV = 0$ (top), and as function of $dV$ for $\epsilon =
    -0.025$
    (bottom).}
    \label{pair2}
\end{center}
\end{figure}
We first analyze the function $F ( \omega )$, which
is a measure of the
on-site pairing. The function $F ( \omega )$ is shown in
Fig.[\ref{pair_onsite1}] as function of the value of $dV$. For
comparison, we plot the same function for $dV = 0$ for different
values of $\epsilon$ in Fig.[\ref{pair_onsite2}]. The two
functions show similar behavior, suggesting that the main effect
of the assisted hopping term on the {\em on-site pairing} correlations is
the renormalization of the level, since its effect
is difficult to be distinguished from the  
effect of a change in the impurity electron level.

 We estimate the tendency towards pairing by using
Eqs.(\ref{pairs}) to define generalized susceptibilities:
\begin{eqnarray}
{\cal F} &= &\int_0^\infty d \omega \frac{F ( \omega ) + F ( -
\omega )}{\omega} \nonumber \\
{\cal F}' &= &\int_0^\infty d \omega \frac{F' ( \omega ) + F' ( -
\omega )}{\omega} \label{susc}
\end{eqnarray}
These functions measure the tendency towards the different types
of pairing. Fig.[\ref{pair1}] gives ${\cal F}$ as function of
$\epsilon$ for $dV = 0$, and as function of $dV$ for $\epsilon = -
0.025$. As it can be appreciated, the inclusion of the assisted
hopping term $dV$ does not change significantly the value of this
susceptibility. Moreover, its magnitude is not too different from
the value obtained for $dV = 0$. On the other hand,
Fig.[\ref{pair2}] gives ${\cal F}'$ as function of $\epsilon$ for
$dV = 0$, and as function of $dV$ for $\epsilon = - 0.025$. In this
case, the effects of $dV$ are quite significant. While of $dV = 0$
this susceptibility remains close to zero for the whole range of
values os $\epsilon$ studied, it raises swiftly as soon as $dV$ is
finite. Hence, the assisted hopping term induces an off diagonal
pairing term of the type described by ${\cal F}'$ given in
Eq.(\ref{pairs}).

This is our main result. In contrast to the {\em local} pairing on
the $d$-level --which can be understood as the consequence of the
primary effect, the renormalization of the level-- the off
diagonal pairing correlations are increased by the enhanced
hopping rate which allows stronger fluctuations if the spin of the
$d$ electron and the local conduction electron have opposite
orientations. There is another interesting feature of this result:
Though the formation of the local moment is strongly suppressed
and the impurity has no well defined spin state, the off diagonal
correlations are still enhanced. This would indicate that this
type of pairing manifests itself in a dynamical way: The spin
state of both the local level and the conduction electron state
fluctuate, but in a correlated way.

\section{Conclusions.}

We have studied, using the Numerical Renormalization Group
method, the Anderson impurity model with a hopping which depends
on the charge state of the impurity. This is the simplest model
which includes information about the internal structure of the
impurity, beyond a single, rigid, electronic state.

Our results indicate that, when the assisted hopping is
sufficiently strong, the model shows a crossover to a phase with
off diagonal pairing correlations. It would be interesting to know
if this regime could be realized in mesoscopic devices, where the
internal structure of a quantum dot can be important.

\section{Acknowledgements.}
We are thankful to J. E. Hirsch and to T. Stauber for a critical
reading of the manuscript and for helpful comments. Financial
support from MCyT (Spain), through grant no. MAT2002-04095-C02-01
is gratefully acknowledged. L.B. acknowledges the financial
support provided through the European Community's Research
Training Networks Programme under contract HPRN-CT-2002-00302,
Spintronics and Hungarian Grants (OTKA) No. T046303, T034243 and
T038162.

\bibliography{NRG}

\end{document}